\documentclass{article}

\usepackage{arxiv}

\usepackage[utf8]{inputenc} % allow utf-8 input
\usepackage[T1]{fontenc}    % use 8-bit T1 fonts
\usepackage{hyperref}       % hyperlinks
\usepackage{url}            % simple URL typesetting
\usepackage{booktabs}       % professional-quality tables
\usepackage{amsfonts}       % blackboard math symbols
\usepackage{nicefrac}       % compact symbols for 1/2, etc.
\usepackage{microtype}      % microtypography
\usepackage{lipsum}
\usepackage{amsmath}
\usepackage{caption} % For caption spacing
\usepackage{subcaption} % For sub-figures
\usepackage{graphicx}
\usepackage{pgfplots}
\usepackage[all]{nowidow}
\usepackage[utf8]{inputenc}
\usepackage{tikz}
\usetikzlibrary{er,positioning,bayesnet}
\usepackage{multicol}
\usepackage{algpseudocode,algorithm,algorithmicx}
\usepackage[inline]{enumitem}

\newcommand{\card}[1]{\left\vert{#1}\right\vert}
\newcommand*\Let[2]{\State #1 $\gets$ #2}
\definecolor{blue}{HTML}{1F77B4}
\definecolor{orange}{HTML}{FF7F0E}
\definecolor{green}{HTML}{2CA02C}

\pgfplotsset{compat=1.14}

\title{An Approach Based on Bayesian Networks for Query Selectivity Estimation}

\author{
  Max Halford \\
  IRIT Laboratory \\
  IMT Laboratory \\
  University of Toulouse \\
  \texttt{max.halford@irit.fr} \\
   \And
 Philippe Saint-Pierre \\
  IMT Laboratory\\
  University of Toulouse \\
  \texttt{philippe.saint-pierre@math.univ-toulouse.fr} \\
   \And
 Frank Morvan \\
  IRIT Laboratory \\
  University of Toulouse \\
  \texttt{frank.morvan@irit.fr} \\
}

\begin{document}
\maketitle

\begin{abstract}
The efficiency of a query execution plan depends on the accuracy of the selectivity estimates given to the query optimiser by the cost model. The cost model makes simplifying assumptions in order to produce said estimates in a timely manner. These assumptions lead to selectivity estimation errors that have dramatic effects on the quality of the resulting query execution plans. A convenient assumption that is ubiquitous among current cost models is to assume that attributes are independent with each other. However, it ignores potential correlations which can have a huge negative impact on the accuracy of the cost model. In this paper we attempt to relax the attribute value independence assumption without unreasonably deteriorating the accuracy of the cost model. We propose a novel approach based on a particular type of Bayesian networks called Chow-Liu trees to approximate the distribution of attribute values inside each relation of a database. Our results on the TPC-DS benchmark show that our method is an order of magnitude more precise than other approaches whilst remaining reasonably efficient in terms of time and space.
\end{abstract}

% keywords can be removed

\section{Introduction}

During query processing \cite{selinger1979access}, each query goes through an optimisation phase followed by an execution phase. The objective of the optimisation phase is to produce an efficient query execution plan in a very short amount of time. The query optimiser draws on the cardinality estimates produced by the cost model for each relational operator in a given plan. Bad cardinality estimates propagate exponentially and have dramatic effects on query execution time \cite{ioannidis1991propagation}. Cardinality estimates are usually made based on a set of statistics collected from the relations and stored in the database's metadata. Such statistics are kept simple in order to satisfy the limited time budget the query optimiser is allocated. However they usually don't capture attribute dependencies.

Formally, given a query $\mathcal{Q}(\mathcal{R}, \mathcal{J}, \mathcal{A})$ over a set of relations $\mathcal{R}$, a set of join predicates $\mathcal{J}$ and a set of attribute predicates $\mathcal{A}$, the cardinality of the query is computed as follows: 

\begin{equation}
    \card{\mathcal{Q}(\mathcal{R}, \mathcal{J}, \mathcal{A})} =  P(\mathcal{J}, \mathcal{A}) \times \prod_{R \in \mathcal{R}} \card{R}
\end{equation}

where $P(\mathcal{J}, \mathcal{A})$ is the selectivity of the query and $\prod_{R \in \mathcal{R}} \card{R}$ is the number of tuples in the Cartesian product of the involved relations. The problem is that $P(\mathcal{J}, \mathcal{A})$ is not available. Moreover estimating it quickly leads to a combinatorial explosion. Simplifying assumptions are made in order to approximate the selectivity whilst ensuring a realistic computational complexity \cite{selinger1979access}.

The first assumption that is commonly made is that attributes are independent within and between each relation. This is the so-called \emph{attribute value independence} (AVI) assumption. It allows to simplify the computation as follows:

\begin{equation}
    P(\mathcal{A}) \simeq \prod_{A_R \in \mathcal{A}} P(A_R) \simeq \prod_{A_R \in \mathcal{A}} \prod_{a_i \in A_R} P(a_i)
\end{equation}

where $P(A_R)$ refers to the selectivity concerning relation $R$ whilst $P(a_i)$ stands for the selectivity of a predicate on attribute $a_i$. In practice the AVI assumption is very error-prone because attributes often exhibit correlations. However it is extremely practical because each distribution $P(a_i)$ can be condensed into a one-dimensional histogram $\widetilde{P}(a_i)$.

Next, the \emph{join predicate independence} assumption implies that join selectivities can be computed independently, which leads to the following approximation:

\begin{equation}
    P(\mathcal{J}) \simeq \prod_{J_i \in \mathcal{J}} P(J_i)
\end{equation}

Assume we are given two relations $R$ and $S$. We want to join both relations on their respective attributes $R.K$ and $S.F$. In this case the selectivity of the join (denoted $J$) can be computed exactly \cite{selinger1979access}:

\begin{equation}
    P(J) = min(\frac{1}{\card{J.R.K}}, \frac{1}{\card{J.S.F}}) 
\end{equation}

The previous assumption doesn't usually hold if multiple foreign keys are included in a join. \cite{ioannidis1991propagation}. Finally, the \emph{join uniformity} assumption states that attributes preserve their distributions after joins. This allows the following simplification:

\begin{equation}
    P(\mathcal{J}, \mathcal{A}) \simeq P(\mathcal{J}) \times P(\mathcal{A})
\end{equation}

Most relational databases \cite{hellerstein2019looking,traverso2013presto,armbrust2015spark} assume all the previous assumptions in conjunction, which leads to the following formula:

\begin{equation}
   P(\mathcal{J}, \mathcal{A}) \simeq \prod_{J_i \in \mathcal{J}} min(\frac{1}{\card{J_i.R.K}}, \frac{1}{\card{J_i.S.F}})  \times \prod_{A_R \in \mathcal{A}} \prod_{a_i \in \mathcal{A_R}} \widetilde{P}(a_i)
\end{equation}

In practice the previous approximation is much too coarse and is frequently wrong by orders of magnitude. However it only requires storage space that grows linearly with the number of attributes and doesn't involve any prohibitive computation. In other words accurate cardinality estimation is traded in exchange for a low computational complexity. The natural question is if a better trade-off is possible. That is, one that relaxes any of the previous assumptions.

A lot of work has gone into developing \emph{attribute-level} synopses \cite{poosala1996improved,heimel2015self} which approximate the distribution $P(a)$ of each attribute $a$. Mostly this involves using histograms and other well-studied statistical constructs. Although theoretically sound, these methods do not help in handling commonplace queries that involve more than one attribute predicate. Furthermore, \emph{table-level} synopses \cite{muralikrishna1988equi} have been proposed to capture dependencies between attributes. The problem is that methods of this kind, such as multi-dimensional histograms, usually require an amount of storage space that grows exponentially with the number of attributes. Table-level synopses also includes various sampling methods \cite{piatetsky1984accurate,lipton1990query,vengerov2015join} where the idea is simply to execute a query on a sample of the database and extrapolate the cardinality. Although they don't handle dependencies across relations, they are computationally efficient because they don't require joins. Finally, \emph{schema-level} synopses \cite{vengerov2015join,leis2017cardinality,chen2017two,kipf2018learned} attempt to soften the join uniformity and join predicate independence assumptions. Although these methods have the potential to handle join-crossing correlations \cite{leis2015good}, they require a prohibitive amount of computational resources because of the amount of joins they necessitate.

Accurate schema-level methods based on Bayesian networks have been proposed \cite{getoor2001selectivity,tzoumas2011lightweight}. A Bayesian network factorises a distribution in order to represent it with a product of lower dimensional distributions. Each lower dimensional distribution captures a dependency between two or more attributes. For example the distribution $P(hair, nationality)$ can be factorised as $P(hair | nationality) \times P(nationality)$ because a person's hair colour is correlated with his nationality. The trick is that finding the right factorisation is an NP-hard problem \cite{jensen1996introduction}. Moreover the time required to produce estimates increases with the complexity of the factorisation \cite{robertson1986graph}. The method proposed in \cite{getoor2001selectivity} successfully captures attribute dependencies across relations but it requires a prohibitive amount of computational complexity that makes it unusable in practice. \cite{tzoumas2011lightweight} propose a simpler method that only attempts to capture dependencies between two relations at most. Although their proposal is more efficient, it still requires performing a significant amount of joins. Moreover the factorisation structures used in both proposals incur an inference procedure that doesn't run in linear time. We believe that giving up some of the accuracy of existing proposals leads to methods that strike a better balance between accuracy and computational complexity. To this extent we propose to factorise the distribution of attributes only inside each relation. We argue that having reliable selectivity estimates for single relations is fundamental for estimating the size of joins \cite{ioannidis1991propagation}. Furthermore we propose to extend a particular type of Bayesian networks called Chow-Liu trees. These allow us to use network structures that are efficient space-wise and can be queried in sub-linear time. Although our approach doesn't capture as many dependencies as in \cite{getoor2001selectivity} and \cite{tzoumas2011lightweight}, it can be compiled quicker and can produce selectivity estimates in less time. Moreover it is still an order of magnitude more precise than trivial models that assume independence. 

The rest of this paper is organised as follows. Section 2 gives an overview of existing methods and their associated pros and cons. This is also where we introduce some notions relating to Bayesian networks. Section 3 is where we describe our model and show how it can efficiently be used for the task of selectivity estimation. Section 4 compares our model to PostgreSQL's cost engine and to a Bernoulli sampling estimator on the TPC-DS benchmark. We explain in what cases our model succeeds and in what cases it doesn't bring anything to the table. Finally, Section 5 concludes and points to some research opportunities.

\section{Related work}

\subsection{Distribution estimation} \label{subsec:statistical-summaries}

The most prominent approach in cost-based query optimisation is to approximate the distribution of attributes of a given database. This has been an area of research ever since \emph{equi-width histograms} were used for summarising a single attribute \cite{kooi1980optimization}. \emph{Equi-height histograms} are commonly used because of their provably lower average error \cite{piatetsky1984accurate}. Meanwhile \cite{ioannidis1993optimal} showed that histograms that minimise the average selectivity estimation error are ones that minimise variance inside each bucket. These histograms are usually called \emph{V-optimal histograms} and involve a prohibitive mathematical optimisation process. As a compromise, \cite{ioannidis1993optimal} introduced the notion of \emph{biased histograms} to find a balance between memorising exact frequencies and approximating them. Histograms are well understood in theory and ubiquitously used in practice, however they don't capture dependencies between attributes.

Multi-column histograms \cite{muralikrishna1988equi,bruno2001stholes,heimel2015self} have been proposed to handle dependencies between two or more attributes. Although they are sound in theory, in practice they are difficult to build and even more so to update \cite{muthukrishnan1999rectangular}. Moreover they require storage space that grows exponentially with the number of attributes.

%As an alternative to histograms, kernel density estimation (KDE) has also been used for %selectivity estimation \cite{blohsfeld1999comparison,heimel2015self}. It can be seen as a smooth %version of a histogram. Although it has good properties and can be used for multi-dimensional %cases, it requires a bandwidth parameter that depends on the data. However the biggest problem %is that KDE can't handle discrete attributes.

To mitigate the exponential growth problem of multi-dimensional histograms, one approach is use a factorised representation of a distribution. The idea is to represent a distribution $P(A_i, \dots, A_n)$ as a product of smaller conditional distributions $P(A_i | Parents(A_i))$. For example the distribution $P(A_1, A_2, A_3)$ can be estimated as $\hat{P}(A_1, A_2, A_3) = P(A_1 | A_2) P(A_3 | A_2) P(A_3)$. $\hat{P}(A_1, A_2, A_3)$ is necessarily an approximation because it doesn't capture the three-way interaction between $A_1$, $A_2$, and $A_3$. The benefit is that although $\hat{P}(A_1, A_2, A_3)$ is an approximation, it requires less storage space. Moreover if $A_1$ and $A_3$ are independent then no information is lost. Bayesian networks \cite{jensen1996introduction} have been shown to be a strong method to find such approximations. However, querying a BN is an NP-hard problem \cite{cooper1990computational} and can take a prohibitive amount of time depending on the structure of the network. Moreover, off-the-shelf implementations don't restrict the structure of the final approximation. This leads to approximations which either require a prohibitive amount of storage space, or are too slow, or both. BNs are classically studied in the context of a single tabular dataset. However in a relational database the data is contained in multiple relations that share relationships. \cite{getoor2001selectivity} first introduced \emph{probabilistic relational models} that could handle the relational setting. They introduced the notion of a \emph{join indicator} to relax the join uniformity assumption. However for structure learning and inference they use off-the-shelf algorithms with running complexities that are way too prohibitive for a database context. \cite{tzoumas2011lightweight} extended this work and proposed to restrict the dependencies a BN can capture to be between two relations at most. Even though their procedure is more efficient, it still requires joining relations, albeit only two at once. Both of these proposals work at a schema-level and require performing a prohibitive amount of joins. Although existing methods based on Bayesian networks seem promising, we argue that they are still too complex to be used at a large scale.

The problem of learning distributions is that they inescapably require a lot of storage space. A radically different approach that has made it's mark is to execute the query on a sample of the database in order to extrapolate the query's cardinality.

\subsection{Sampling} \label{sec:sampling}

Sampling is most commonly used to estimate selectivity for a query that pertains to a single relation \cite{piatetsky1984accurate}. The simplest method is to sample a relation $R_i$ with probability $p_i$. The obtained sample $r_i$ will then contain $\card{R_i} \times p_i$ tuples. To estimate the selectivity of a query on $R_i$ one may run the query on it's associated sample $r_i$ and multiply the cardinality of the output by $\frac{1}{p}$. This is commonly referred to as \emph{Bernoulli sampling} and works rather well given a sufficient sample size. Adaptive methods \cite{lipton1990query} have also been proposed to determine an optimal sample size for each relation. Sampling is attractive because it is simple to implement and naturally captures dependencies between attributes. Moreover, sampling can be performed on multiple relations in order to capture inter-relational attribute dependencies.

%zhao2018random in third cite

Sampling across multiple relations is a difficult task. Indeed \cite{chaudhuri1999random} showed that the join of independent uniform samples of relations is not a uniform sample of the join of the relations. Many methods have been proposed for two-way joins \cite{olken1993random,vengerov2015join}. Their common idea is to use a hash function $h(a) \rightarrow [0, 1]$ to make sure joined samples share keys. Say $R_1$ has an attribute $A_1$ which is a foreign key to an attribute $A_2$ of a relation $R_2$. By applying the same hash function $h(a)$ to both attributes, one may obtain samples which preserve join relationships by keeping all the tuples that satisfy $h(a) < p$. This way all tuples from both relations that satisfy $h(a) < p$ will be included in the sample. The unbiased estimator for the size of the result of the join is $J(R_1, R_2)$ is $\frac{J(r_1, r_2)}{p}$. Although quite strong in theory, join-aware sampling \cite{li2016wander} requires a prohibitive full pass over the involved relations if no index is available. Moreover, this approach doesn't necessarily extend to joins involving more than two relations \cite{acharya1999join}.

\subsection{Learning}

To completely sidestep the difficulties inherent to query optimisation, learning procedures that correct their mistakes have been proposed \cite{stillger2001leo,chen1994adaptive,kipf2018learned}. In the case of database optimisation learning has been used to tune various models and to memorise observed selectivities. This is done by \emph{query feedback} where the cost model gets access to the actual cardinalities \cite{chen1994adaptive} after the query execution phase. By comparing the estimates it has made with the actual values it can make adjustments with the goal of making less mistakes for subsequent queries. The most successful method in this category is DB2's LEO optimiser \cite{stillger2001leo}. The approach LEO takes is simply to memorise the true cardinality of error-prone parts of executed query plans. This works remarkably well in an environment where a given query is run repeatedly. However it doesn't help for estimating the cardinality of unseen queries. Recently an interesting approach based on deep learning has also been proposed \cite{kipf2018learned}. Apart from LEO, learning approaches have not yet matured enough to be used in practice.

\subsection{Discussion}

All of the previously mentioned methods offer a different compromise in terms of accurate cardinality estimation, temporal complexity, and spatial complexity. On the one hand, histograms are precise, quick, and lightweight. However, they require a prohibitive amount of storage space if one wishes to capture attribute dependencies. On the other hand, sampling can easily capture attribute dependencies; but it is too slow because either the sample has to be constructed online or loaded in memory. Finally learning is an original take on the problem but it doesn't help for unseen queries. Our contribution is to use Bayesian networks to factorise the distribution of the attributes of each relation. This way we capture the most important attribute dependencies and ignore the unimportant ones to preserve storage space. A huge benefit of our method is that we can optimise each Bayesian network on a sample of the associated relation to save time without a significant loss in accuracy. The downside is that like most methods we ignore dependencies between attributes of different relations.

\section{Methodology}

\subsection{Finding a good network} \label{sec:structure-learning}

\begin{figure}[H]
\begin{subfigure}{.5\textwidth}
  \centering
        \begin{tikzpicture}[shorten >= 1pt, auto, thick]
        
          % Define nodes
          \node[latent] (Nationality) {N};
          \node[latent, below=0.5 of Nationality, xshift=-1cm] (Hair) {H};
          \node[latent, below=0.5 of Nationality, xshift=1cm] (Gender) {G};
        
          % Connect the nodes
          \edge {Nationality} {Hair}
          \edge {Nationality} {Gender}
        
        \end{tikzpicture}
\end{subfigure}%
\begin{subfigure}{.5\textwidth}
  \centering
   \begin{tikzpicture}[shorten >= 1pt, auto, thick]
        
          % Define nodes
          \node[latent] (Hair) {H};
          \node[latent, above=0.5 of Hair, xshift=-1cm] (Nationality) {N};
          \node[latent, above=0.5 of Hair, xshift=1cm] (Gender) {G};
        
          % Connect the nodes
          \edge {Nationality} {Hair}
          \edge {Gender} {Hair}
        
        \end{tikzpicture}
\end{subfigure}
\caption{Possible factorisations of $P(hair, nationality, gender)$}
\label{fig:example-networks}
\end{figure}
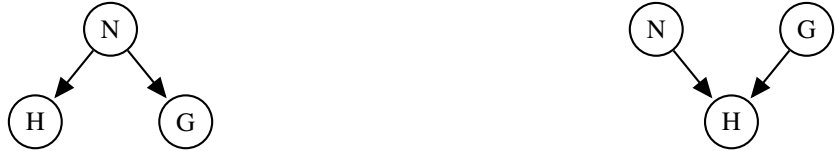

A Bayesian network (BN) factorises a probability distribution $P(X_1, \dots, X_n)$ into a product of conditional distributions. For any given probability distribution $P(\mathcal{X})$ there exist many possible BNs. For example $P(hair, nationality, gender)$ can be factorised as $P(hair | nationality) P(gender | nationality) P(nationality)$ as well as $P(hair | (nationality, gender))P(nationality)P(gender)$ (see figure \ref{fig:example-networks}).

The goal of \emph{structure learning} is to find a BN that closely matches $P(\mathcal{X})$ whilst preserving a low computational complexity. Indeed, for any given BN, the cost of storing it and of computing a marginal distribution $P(X_i)$ depend on it's structure.

The classic approach to structure learning is to define a scoring function that determines how good a BN is -- both in terms of accuracy and complexity -- and to run a mathematical optimisation procedure over the possible structures \cite{heckerman1995learning}. The problem is that such kind of procedures are too costly and don't fit inside the tight computational budget a database typically imposes. Recently linear programming approaches that require an upper bound on the number of parents have also been proposed \cite{jaakkola2010learning}; in practice these can handle problems with up to 100 variables which is far from ideal. Finally, one can also resort to using greedy algorithms that run in polynomial time but don't necessarily find a global optimum. 

\emph{Chow-Liu trees} \cite{chow1968approximating} is one such method that finds a BN where dependencies between two attributes are the only ones considered. Building a Chow-Liu tree only involves three steps. Initially the mutual information (MI) between each pair of attributes is computed. These values define a fully connected graph $G$ where each MI value is translated to a weighed edge. Next, a minimum spanning tree (MST) of $G$ is retrieved. This can be done in $\mathcal{O}(nlog(n))$ time where $n$ is the number of attributes. Finally the MST has to be directed by choosing a node at random and defining it as the root. 

We choose to use Chow-Liu trees for two practical reasons. First of all they are simple to construct. The only part that doesn't scale well is computing the MI values. However this can be accelerated by using a coarser representation of the data such as histograms. Moreover the process can be run over a sample of a relation. In our experience these two tricks greatly reduced computation time without hindering the accuracy of the resulting trees. Secondly the output network is a tree -- hence there is only one parent per node. This is practical because retrieving a marginal distribution -- in other words \emph{inferring} -- from a tree can be done in linear time \cite{robertson1986graph}. Moreover, storing the network only requires saving $n-1$ two-dimensional distributions and one uni-dimensional distribution. On top of this, \cite{chow1968approximating} proves that Chow-Liu trees minimise the KL divergence, meaning that they are the best possible trees from an information theory perspective. The downside is that they can't capture dependencies between more than 2 variables -- for example it only snows if it's cold and rainy. However in our experience these kind of dependencies are not so common.

\subsection{Estimating the conditional probabilities} \label{sec:parameter-estimation}

Once a satisfying structure has been found, the necessary probability distributions have to be computed. Indeed recall that a Bayesian network is nothing more than a product of conditional probability distributions (CPD). A CPD gives the distribution of a variable given the value of one or more so-called parent variables. For example tables \ref{tab:hair-nationality} and \ref{tab:gender-nationality} are two CPDs that are both conditioned on the \emph{nationality} variable. 

\begin{table}[H]
\centering
\makebox[0pt][c]{\parbox{1.3\textwidth}{%
    \begin{minipage}[t]{0.3\hsize}\centering
        \begin{tabular}{@{}cc@{}}
        \textbf{American}    & \textbf{Swedish} \\ \hline
        0.5 & 0.5 \\
        \end{tabular}
        \caption{$P(nationality)$}
    \end{minipage}
    \begin{minipage}[t]{0.35\hsize}\centering
        \begin{tabular}{@{}c|ccc@{}}
                       & \textbf{Blond} & \textbf{Brown}  & \textbf{Dark} \\ \hline
    \textbf{American}           & 0.2               & 0.6         & 0.2      \\ \hline
    \textbf{Swedish}           & 0.8                  & 0.2         &  0     \\
            \end{tabular}
        \caption{$P(hair | nationality)$}
        \label{tab:hair-nationality}
    \end{minipage}
    \begin{minipage}[t]{0.3\hsize}\centering
        \begin{tabular}{@{}c|cc@{}}
                       & \textbf{Male} & \textbf{Female} \\ \hline
                         \textbf{American}           & 0.5              & 0.5        \\ \hline
\textbf{Swedish}           & 0.45              & 0.55        \\
        \end{tabular}
        \caption{$P(gender | nationality)$}
        \label{tab:gender-nationality}
    \end{minipage}%
}}
\vspace{-4mm}
\end{table}

The number of values needed to define a CPD is $c^{p+1}$ where $c$ is the cardinality of each variable -- for simplicity we assume it is constant -- and $p$ is the number of parent variables. This stems from the fact that each CPD is related to $p+1$ variables and that each and every combination of values has to be accounted for. The fact that Chow-Liu trees limits the number of parents each node has to 1 means that we only have to store $c^2$ values per distribution. Moreover a sparse representation can be used to leverage the fact that 0s are frequent. However, if the cardinality of a variable is high then a lot of values still have to be stored. This can be rather problematic in a constrained environment.

To preserve a low spatial complexity we propose to use end-biased histograms described in subsection \ref{subsec:statistical-summaries}. The idea is to preserve the exact probabilities for the $k$ most common values of a variable and put the rest of the probabilities inside $j$ equi-height intervals. Using equi-height intervals means that we don't have to store the frequency of each interval. Indeed it is simply $1 - \sum_{i=1}P(MCV_i)$ where $P(MCV_i)$ denotes the frequency of the $i^{th}$ most common value. Instead, by assuming that the values inside an interval are uniformly distributed, we only have to store the number of distinct values the interval contains. Table \ref{tab:compressed-hair-nationality} shows what a CPD with intervals looks like. In the example, given that a person is American, there is probability of $1 - (0.2 + 0.5) = 0.3$ that his hair colour is in the [Dark, Red] interval. Because there are 3 distinct values in the [Dark, Red] interval, the probability that an American has, say, hazel hair is $\frac{1 - (0.2 + 0.5)}{3} = 0.1$. 

\begin{table}[H]
\centering
\begin{tabular}{@{}c|ccc@{}}
                                & \textbf{Blond} & \textbf{Brown}   & \textbf{[Dark, Red]} \\ \hline
    \textbf{American}           & 0.2            & 0.5              & 3                   \\ \hline
    \textbf{[British, French]}  & 0.4            & 0.3              & 3                   \\ \hline
    \textbf{Swedish}            & 0.8            & 0.2              & 0                   \\
            \end{tabular}
        \caption{$P(hair | nationality)$ with $k=2$ and $j=1$}
        \label{tab:compressed-hair-nationality}
\vspace{-4mm}
\end{table}

Compressing a CPD this way means we only have to store $(k+j)^2$ values per distribution. If we assume that there are $n$ attributes inside a relation then storing a Bayesian networks requires $(k+j) + (n-1)(k+j)^2$ values in total -- the first $(k+j)$ corresponds to the network's root node which is not conditioned on any other variable. This has the added advantage that we can handle continuous variables that usually have extremely high cardinalities.

Fortunately, retrieving CPDs inside a relational database can easily be done with the \texttt{GROUP BY} and \texttt{COUNT} statements. Moreover, the CPDs can be computed on a sample of the relations to reduce computation time. Whats more, if data is appended to the database then only the CPDs have to recomputed if one assumes the structures of the Bayesian networks remain constant through time. However, if new attributes are added to a relation then the structure of it's Bayesian network has to be rebuilt from scratch.   

\subsection{Producing selectivity estimates} \label{sec:inference}

As previously mentioned, inference is the task of obtaining a marginal distribution from a Bayesian network. For example we may want to know the probability of Swedish people having blond hair (i.e.\ $P(hair=``Blond" \wedge nationality=``Swedish")$). The idea is to treat the obtained probability as the selectivity of the associated relational query. For each relation involved in a query, we identify the part of the query that applies to the relation and determine it's selectivity. Then, by assuming that attributes from different relations are independent, we simply multiply the selectivities together. Although this is a strong assumption, we argue that capturing table-level dependencies can still have a significant impact on the overall cardinality estimation. Of course we would be even more precise if we had determined dependencies between different relations as in \cite{getoor2001selectivity,tzoumas2011lightweight}, but it would necessarily involve joins. In other words our method offers a different trade-off between accuracy and computational feasibility.

Performing inference over a BN is an NP-hard problem \cite{cooper1990computational}. However, because we have restricted our BNs to trees, we can make full use of purpose-built algorithms that only apply to trees. The \emph{variable elimination} (VE) algorithm \cite{cowell2006probabilistic} is a simple exact inference algorithm that can be applied to any kind of network topology. Specifically the complexity of VE is $\mathcal{O}(n\exp{(w)})$ where $n$ is the number of nodes and $w$ is the width of the network \cite{robertson1986graph}. However the width of a tree is necessarily 1, meaning VE can run in $\mathcal{O}(n)$ time. The formula for applying VE is given in \eqref{inference}, wherein $k$ attributes are being queried out of a total of $n$.

\begin{equation}
    \label{inference}
    P(A_1=a_1, \dots, A_k=a_k) = \sum_{i=k+1}^n\prod_{j=1}^{k} P(A_j=a_j | Parents(A_j))
\end{equation}

%For example, if $hair$ depends on $nationality$ and $nationality$ depends on $gender$, then by %applying equation \eqref{inference} we obtain:

%\begin{equation*}
%\begin{split}
%P(hair = ``Blond") & = P(hair = ``Blond" | nationality = ``Swedish") \; + \\
%                   & \quad\; P(hair = ``Blond" | nationality = ``American") \\
%                   & = P(hair = ``Blond" | nationality = ``Swedish") \; \times \\
%                   & \quad\; (P(nationality = ``Swedish" | gender = ``Male") \; + \\
%                   & \quad\; P(nationality = ``Swedish" | gender = ``Female")) \; + \\
%                   & \quad\; P(hair = ``Blond" | nationality = ``American") \; \times \\
%                   & \quad\; (P(nationality = ``American" | gender = ``Male") \; + \\
%                   & \quad\; P(nationality = ``American" | gender = ``Female"))
%\end{split}
%\end{equation*}

The idea of VE is to walk over the tree in a post-order fashion -- i.e.\ start from the leaves -- and sum up each CPD row-wise. This avoids unnecessarily computing sums more than needed and ensures the inference process runs in linear time. The computation can be further increased by noticing that not all nodes in a BN are needed to obtain a given marginal distribution \cite{jensen1996introduction}. Indeed the VE algorithm only has be run on a necessary subset of the tree's nodes which is commonly referred to as the \emph{Steiner tree} \cite{hwang1992steiner}. Extracting a Steiner tree from a tree can be done in linear time (see algorithm \ref{algo:steiner}).

\begin{figure}[H]
\centering
\begin{tikzpicture}

  % Define nodes
  \node[latent, draw=blue!80, line width=0.5mm] (G) {G};
  \node[latent, below=0.5 of G, xshift=-1cm] (S) {S};
  \node[latent, below=0.5 of G, xshift=1cm, draw=blue!80, line width=0.5mm] (N) {N};
  \node[latent, below=0.5 of N, xshift=-1cm, draw=blue!80, line width=0.5mm] (H) {H};
  \node[latent, below=0.5 of N, xshift=1cm] (P) {P};

  % Connect the nodes
  \edge {G} {S} ;
  \edge[blue!80, line width=0.5mm] {G} {N} ;
  \edge[blue!80, line width=0.5mm] {N} {H}
  \edge {N} {P} ;

\end{tikzpicture}
\caption{Steiner tree in blue containing nodes G, N, and H needed to compute H's marginal distribution}
\end{figure}
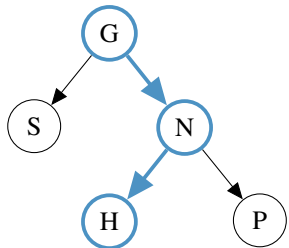

In our case we are using CPDs with intervals, meaning that we have to tailor the VE algorithm around them. Fortunately this is quite simple as we only have to check if a given value is an interval or not. Range queries can be handled by interpolating inside the interval whilst for equality queries we can assume that all distinct values in the interval are equally frequent.

\begin{algorithm}[H]
  \caption{Steiner tree extraction}
  \begin{algorithmic}[1]
    \Function{Walk}{$node, required, path, relevant$}
      \If{required is empty}
        \State \Return{$\{\}$}
      
      \ElsIf{node in nodes}
        \Let{$required$}{$required \setminus \{node\}$}
        \Let{$relevant$}{$relevant \cup path$}
      \EndIf
      
      \Let{$path$}{$path \cup \{node\}$}
      
      \For{$child \in node.children()$}
        \Let{$relevant$}{$relevant \  \cup$ \Call{Walk}{$child, required, path, relevant$}}
      \EndFor
      
      \State \Return{$relevant$}
    \EndFunction
    \Statex
    \Function{ExtractSteinerTree}{$tree, nodes$}
      \Let{$nodes$}{$nodes \cup tree.root()$}
      \Let{$relevant$}{\Call{Walk}{$tree, nodes, \{\}, \{\}$}}
      \State \Return{$tree.subset(relevant))$}
    \EndFunction
  \end{algorithmic}
  \label{algo:steiner}
\end{algorithm}

\section{Experimental study}

\subsection{Setup}

We implemented a prototype of our method along with the Bernoulli sampling described in section \ref{sec:sampling} and the textbook method described in the introduction. We chose these two methods because they are realistic and are used in practice. We would have liked to compare our method to previous Bayesian approaches proposed in \cite{getoor2001selectivity} and \cite{tzoumas2011lightweight}, however we were not able to accurately reproduce their results given the available information. We expect our method to be less accurate but much more computationally efficient. Our goal is to quantitatively show why our method offers a better trade-off than the other two implemented methods. We ran all methods against a small subset of the queries contained in the TPC-DS benchmark \cite{poess2007you} with a scale factor of 20 \footnote{Specifically we used the following queries: 7, 13, 18, 26, 27, 53, 54, 91}. We only picked queries that apply more than one attribute predicate on at least one relation and that exhibit dependencies. We chose this subset on purpose because our model doesn't bring anything new to the table if there is only one predicate. Indeed if there is only one predicate then our model is equivalent to the textbook approach of using one histogram per attribute.

We used four criteria to compare each method: 
\begin{enumerate*}[label=(\arabic*)]
    \item The construction time.
    \item The accuracy of the cardinality estimates.
    \item The time needed to make a cardinality estimate.
    \item The number of values needed to store the model.
\end{enumerate*}
We ran our experiments multiple times to get statistically meaningful results. First of all we used 10 different sample sizes to determine the impact of sampling. Then, for each combination of method and sampling size we took measurements with 10 different seeds. For each measurement we thus calculated it's mean and it's standard deviation. To make the comparison fair we used equi-height histograms with the same parameters for both the textbook and the Bayesian networks approaches. Specifically we stored the exact frequencies of the 30 most common values and approximated the rest with 30 buckets.

\subsection{Construction time}

\begin{figure}[!ht]
    \centering
    \scalebox{0.55}{\input{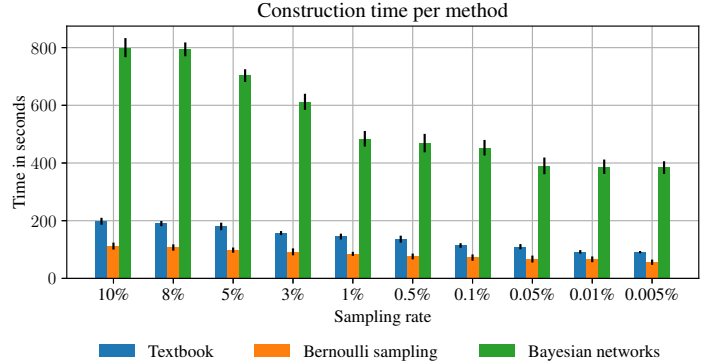}}
    \caption{Construction time}
    \label{fig:construction-time}
\end{figure}

We first of all measured the time it takes to construct each model (see figure \ref{fig:construction-time}). Naturally \textcolor{orange}{sampling} is the method that takes the least time because nothing has to be done once the sample is retrieved from the database. The \textcolor{blue}{textbook} and \textcolor{green}{Bayesian network} (ours, that is) methods necessarily take longer because they have to perform additional computations after having obtained the sample. The \textcolor{blue}{textbook} method only has to build equi-height histograms. The \textcolor{green}{Bayesian network} method requires slightly more involved calculations. It spends most of it's time computing mutual information scores and processing \texttt{GROUP BY} operations. Although these operations are unavoidable, their running time can be mitigated as explained in section \ref{sec:structure-learning}. Moreover the \texttt{GROUP BY} results used for computing mutual information scores can be reused for parameter estimation as explained in section \ref{sec:parameter-estimation}. However for the sake of simplicity we didn't implement these optimisations in our prototype. Still, the results we obtained seem better than those presented in \cite{tzoumas2011lightweight} where the authors claim their method can process a database of 1GB in about an hour. Our method can process the same amount of data in under 8 minutes.

\subsection{Cardinality estimates}

We then compared methods based on their average accuracy. In other words we ran each method against each query and measured the average error. Although our method improves the accuracy of cardinality estimation for queries on a single relation; our goal in this benchmark is to measure how much this will impact the overall accuracy for general queries over multiple relations. It is possible that some queries bias the average accuracy because each queries returns a number of rows that can vary in magnitude in regard to the others. Because of this, typical metrics such as the mean squared error (MSE) can't be used. \cite{moerkotte2009preventing} explain why using average multiplicative errors makes the most sense in the context of query optimisation. For a given number of rows $y$ and an estimate $\hat{y}$ we calculated the $q$-error \cite{leis2018query} which is defined as $q(y, \hat{y}) = \frac{max(y, \hat{y})}{min(y, \hat{y})}$.

\begin{figure}[!ht]
    \centering
    \scalebox{0.55}{\input{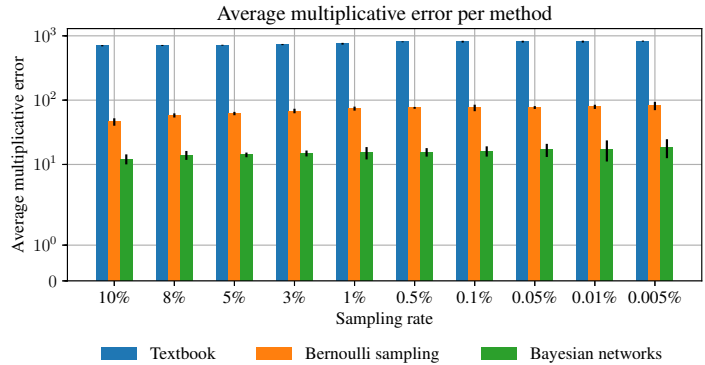}}
    \caption{Average errors}
    \label{fig:errors}
\end{figure}

The advantage of the $q$-error is that it returns an error that is independent of the scale of the values at hand. Moreover the $q$-error is symmetric and will thus be the same regardless of the fact that we are underestimating or overestimating the cardinality. For each combination of method and sampling rate we averaged the $q$-error over all 8 queries. As previously mentioned we took measurements with different samples so to reduce any possible bias in the results. The results are displayed in figure \ref{fig:errors}.

Unsurprisingly, the \textcolor{blue}{textbook} method produces estimates that are off by several orders of magnitude. What is more interesting is that \textcolor{green}{Bayesian networks} are significantly better than \textcolor{orange}{sampling}. The reason this occurs is because the sampling method doesn't place any uncertainty as to if a value is present in a relation or not. A value is either in a sample or not. Meanwhile the Bayesian networks method uses histograms to approximate the frequencies of the least common values. This has a significant impact on the overall average, at least for the subset of queries we chose.

\subsection{Inference time}

\begin{figure}[!ht]
    \centering
    \scalebox{0.55}{\input{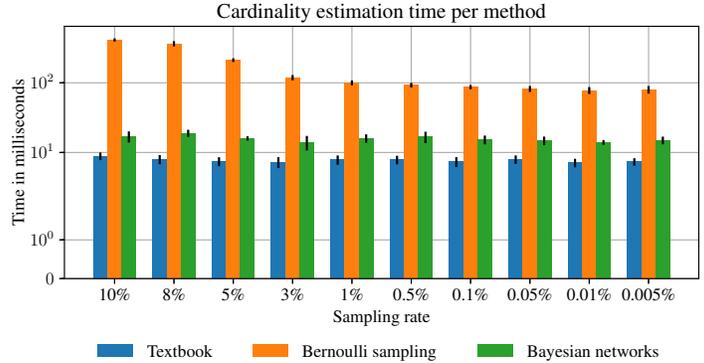}}
    \caption{Cardinality estimation time}
    \label{fig:estimation-time}
\end{figure}

We also measured the average time it takes to estimate the selectivity of a query. A query optimiser has a very small amount of time to produce a query execution plan. Only a fraction of this time can be allocated to cardinality estimation. It is thus extremely important to produce somewhat accurate cardinality estimates in a timely fashion. We recorded our results in figure \ref{fig:estimation-time}.

As can be seen the main pitfall of the \textcolor{orange}{sampling} method is that it takes a considerable amount of time to estimate a cardinality. This is expected because for each set of predicates a full pass has to be made on the according sample. Whats more we didn't even take into account the fact that the necessary samples have to be loaded in memory beforehand. As for the \textcolor{blue}{textbook} method, it only has to read values from a histogram. Meanwhile the \textcolor{green}{Bayesian networks} method has to extract the Steiner tree and perform variable elimination on it as explained in section \ref{sec:inference}. This is naturally more burdensome than simply looking up values in a histogram, but it is still on the same order of magnitude in terms of time.

\subsection{Disk usage}

Finally we measured the number of values needed to store each model. For the sampling method each and every sample has to be stored. Meanwhile the textbook and Bayesian networks methods are synopses and require storing a significantly lesser amount of values. In our experiments the worse case storage bounds of both of these methods are pessimistic. For example in our experiments, the theoretical upper bound textbook method is around 32000 values, but only \%53 of the values actually need to be stored (the other 47\% are 0s). Moreover the same occurs for the Bayesian networks method; indeed for a 5\% sample only around 300000 values out of the theoretical 400000 have to be stored. This is due to the fact that some attributes have a very low number of unique values which makes the associated histograms smaller than expected.

\begin{table}[H]
\centering
\begin{tabular}{@{}c|ccc@{}}
                       & \textbf{Size} & \textbf{Sparsity} & \textbf{Effective size} \\ \hline
\textbf{Textbook}           & 117KB              & 47\%   & 62KB     \\ \hline
\textbf{Sampling}           & 412MB              & 0\%      & 412MB  \\ \hline
\textbf{Bayesian network}           & 615KB      & 24\%         & 467.4KB       \\
        \end{tabular}
        \caption{Storage size per method using a 5\% sampling rate}
        \label{tab:store-size}
\vspace{-4mm}
\end{table}

The numbers presented in table \ref{tab:store-size} were obtained by using a 5\% sample of the database. Apart from the sampling method the numbers are more or less the same when using different sample sizes. Indeed histograms and conditional probability distributions have a fixed size which doesn't increase with the amount of data they synthesise. Meanwhile using sampling means that the whole has to be stored either in memory or on the disk. We noticed that the higher the dependencies between the attributes, the higher the sparsity of the conditional probability distributions. This is expected because of soft functional dependencies that lead the conditioned histograms to possess only a few values. 

\section{Conclusion}

The majority of cost models are blindfolded and do not take into account attribute dependencies. This leads to cardinality estimation errors that grow exponentially and have a negative impact on the query execution time. To prevent this we propose a novel approach based on Bayesian networks to relax the independence assumption. In contrast to prior work also based on Bayesian networks we only capture dependencies inside each relation. This allows our method to be compiled in much less time and to produce selectivity estimates in sub-linear time. We do so by restricting the structure of the network to a tree and by compressing each attribute's conditional probability distributions. We ran our method on a chosen subset of the TPC-DS benchmark and obtained satisfying results. Our method is an order of magnitude more accurate than estimates that assume independence, even though it doesn't attempt to capture cross-relational dependencies. Although our method requires storing a few two-dimensional distributions, the storage requirements are a tiny fraction of those of sampling methods.

Like other table-level synopses, our model does not capture dependencies between attributes of different relations. Whats more it doesn't help in determining the size of multi-way joins. In the future we plan to work on these two aspects of the cardinality estimation problem.

\bibliographystyle{unsrt}  


\end{document}